\documentclass[11pt,letterpaper]{article}
\renewcommand{\footnoterule}{%
  \kern -3pt
  \hrule width \textwidth height 0.5pt
  \kern 2pt
}
\usepackage{latexsym}
\usepackage{amssymb}
\usepackage{amsmath}
\usepackage{graphicx}
\usepackage{bm}
\usepackage{enumerate}
\usepackage{sectsty}
\usepackage{subfig}
\usepackage{float}
\usepackage{caption}
\usepackage{times,fancyhdr}
\usepackage{dsfont}
\usepackage{wasysym}
\sectionfont{\normalsize}
\subsectionfont{\small}

\newcommand{\romansubs}{\renewcommand{\theequation}{\theparentequation \roman{equation}}}

\newcommand{\hr}{{h^{a}_{\mbox{\tiny{(r)}}\mu}}}
\newcommand{\neut}{PSR J0045-7319 }
\newcommand{\msun}{M_{\mbox{\tiny{$\astrosun$}}}}

\newcommand*{\Scale}[2][4]{\scalebox{#1}{$#2$}}%

\begin{document}

\pagestyle{fancy}
\fancyhead{} % clear all header fields
\fancyhead[OR]{\thepage}
\fancyhead[OC]{{\small{THE SPHERICALLY SYMMETRIC VACUUM IN COVARIANT $F(T)$ GRAVITY}}}
\fancyfoot{} % clear all footer fields
\renewcommand\headrulewidth{0.5pt}
\addtolength{\headheight}{2pt} % make space for the rule

\title{{\small \bf{THE SPHERICALLY SYMMETRIC VACUUM IN COVARIANT $\bm{F(T)}$$\bm{=T+\frac{\alpha}{2}T^{2}+\mathcal{O}(T^{\gamma})}$ GRAVITY THEORY}}}
\date{}
\maketitle
\begin{center}
\vspace{-1.5cm}{{\small{\;\;}}\;\;\;{\small{Andrew DeBenedictis}$^\dagger$}\;\;\;{\small{Sa\v{s}a Iliji\'{c}}$^\ddagger$}\;\;\;{\small{}}}
\end{center}
{\let\thefootnote\relax\footnote{\begin{tabular}{ l l l l}
  $\dagger${\scriptsize{The Department of Physics, and}} & \;$\ddagger${\scriptsize{Department of Applied Physics,}} & \;{\scriptsize{}} & \;{\scriptsize{}} \\
  \;\,{\scriptsize{The Pacific Institute for the Mathematical Sciences,}} & \;\;\,{\scriptsize{Faculty of Electrical Engineering and Computing,}} & \;\;\,{\scriptsize{}} & \;\;\,{\scriptsize{}}\\
  \;\,{\scriptsize{Simon Fraser University,}} & \;\;\,{\scriptsize{University of Zagreb,}} & \;\;\,{\scriptsize{}} & \;\;\,{\scriptsize{}}\\
  \;\,{\scriptsize{Burnaby, British Columbia, Canada}} & \;\;\,{\scriptsize{Zagreb, Croatia}} & \;\;\,{\scriptsize{}} & \;\;\,{\scriptsize{}} \\
  \;\,{\scriptsize{adebened@sfu.ca}} & \;\;\,{\scriptsize{sasa.ilijic@fer.hr}} & \;\;\,{\scriptsize{}} & \;\;\,{\scriptsize{}}\\
  \;\,{\scriptsize{}} & \;\;\,{\scriptsize{}} & \;\;\,{\scriptsize{}} & \;\;\,{\scriptsize{}}
\end{tabular}}}

\setcounter{footnote}{0}
\begin{abstract}
\noindent Recently, a fully covariant version of the theory of $F(T)$ torsion gravity has been introduced \cite{ref:covariant}. In covariant $F(T)$ gravity the Schwarzschild solution is not a vacuum solution for $F(T)\neq T$ and therefore determining the spherically symmetric vacuum is an important open problem. Within the covariant framework we perturbatively solve the spherically symmetric vacuum gravitational equations around the Schwarzschild solution for the scenario with $F(T)=T + (\alpha/2)\, T^{2}$, representing the dominant terms in theories governed by Lagrangians analytic in the torsion scalar. From this we compute the perihelion shift correction to solar system planetary orbits as well as perturbative gravitational effects near neutron stars. This allows us to set an upper bound on the magnitude of the coupling constant, $\alpha$, which governs deviations from General Relativity. We find the bound on this nonlinear torsion coupling constant by specifically considering the uncertainty in the perihelion shift of Mercury. We also analyze a bound from a similar comparison with the periastron orbit of the binary pulsar \neut as an independent check for consistency. Setting bounds on the dominant nonlinear coupling is important in determining if other effects in the solar system or greater universe could be attributable to nonlinear torsion.
\end{abstract}
\rule{\linewidth}{0.2mm}
\vspace{-1mm}

\section{{Introduction}}
The extended teleparallel gravity is a theory of gravity which is based purely on torsion, as opposed to curvature as in general relativity. In this theory the fundamental field is not the metric but instead is the tetrad. The teleparallel equivalent of general relativity (TEGR) is the one where the Lagrangian density is linear in the torsion scalar whereas the extended teleparallel gravity's Lagrangian density also possesses nonlinear terms. The action for the extended theory can be written as\footnote{Latin indices are orthonormal Lorentz indices whereas Greek indices are spacetime indices.}
\begin{equation}
S=\int \left\{\frac{F(T)}{2\kappa}+\mathcal{L}_{\mbox{\tiny{m}}}\right\}\,\mbox{det}[h^{a}_{\;\mu}]\,d^{4}x\,, \label{eq:gravaction}
\end{equation}
where $\mathcal{L}_{\mbox{\tiny{m}}}$ represents the matter Lagrangian density, $\kappa=8\pi$, $h^{a}_{\;\mu}$ is the tetrad, and $F(T)$ is a function of the torsion scalar, $T$. When $F(T)$ is linear in $T$, and the matter coupling is minimal \cite{ref:DandS}, \cite{ref:DandS2}, the theory is equivalent to general relativity. However, when $F(T)$ possesses terms non-linear in $T$ the theory can be radically different from the corresponding $F(R)$ theory. One interesting property of $F(T)$ theory not shared by $F(R)$ theory is that the resulting field equations remain second-order regardless of the form of $F(T)$. 

There has been much interesting work done in $F(T)$ theory, mainly in the realm of cosmology (see \cite{ref:cosmo1} - \cite{ref:cosmoend} and references therein). To a lesser extent, black holes have also been studied in $F(T)$ theory \cite{ref:bh1} - \cite{ref:bhend}. One issue that has been problematic with $F(T)$ gravity is that, in the usual formulation, the theory is not locally Lorentz invariant when $F(T)\neq T$. Therefore, inequivalent equations of motion are obtained by Lorentz transforming the tetrad field. Because of this, researchers were forced into searching for what are known as ``good'' (vs ``bad'') tetrads \cite{ref:goodbadtets}. It was not always clear what constituted a good tetrad. In general, one could appeal to the following criteria \cite{ref:goodbadtets},\cite{ref:bh2}:
\begin{itemize}\label{page:criteria}
\item The tetrad should not place restrictions on the form of $F(T)$. That is, the tetrad needs to retain acceptable equations of motion regardless of the function $F(T)$ and not just work well for certain functions. In the past the following conditions were proposed for acceptable equations of motion:
\item The tetrad must produce equations of motion which are compatible with a symmetric stress-energy tensor $\mathcal{T}_{\mu\nu}=\mathcal{T}_{\nu\mu}$.
\item The resulting equations of motion should not produce peculiar physics. For example, in spherical symmetry there should be no energy transport in the angular directions. In a static scenario, there should be no energy flux from one location to the other, etc.
\end{itemize}
It was realized that the reason for the lack of local Lorentz invariance in the non covariant theory was related to the fact that the theory did not contain simply the gravitational degrees of freedom, but also ones which depend on the particular Lorentz frame one is in \cite{ref:earlycov}, \cite{ref:earlycov2}, \cite{ref:davood}. This comes in via the inertial part of the spin connection\footnote{We slightly abuse nomenclature in that ``spin connection'' always refers to $\omega$ and we refer to the transport connection of the theory explicitly as the Weitzenb\"{o}ck connection regardless of index character.}  
\begin{equation}
\omega^{a}_{\;\;b \sigma}=\Lambda^{a}_{\;\;c}\partial_{\sigma}\Lambda_{b}^{\;\;\;c}, \label{eq:spincon}
\end{equation}
where $\Lambda^{\cdot}_{\;\cdot}$ are the usual Lorentz transformation matrix components. Note that the purely inertial spin connection above depends only on which Lorentz frame and it can be shown to yield no curvature. It can therefore contribute to curvatureless torsion-only theory such as $F(T)$ gravity, and its ambiguity (as there are infinitely many valid Lorentz frames) is a source of introducing inertial effects in non covariant $F(T)$ theory. Therefore,
in the non-covariant theory one must choose a tetrad where a {purely inertial} spin connection (the specific spin connection we are interested in will be discussed below) vanishes, for the tetrad to be a truly good tetrad. In this particular (non covariant) case the connection is then the curvatureless Weitzenb\"{o}ck connection, $\Gamma^{\alpha}_{\;\;\beta\gamma}$, but without the spin connection, and the torsion tensor, $T^{\alpha}_{\;\;\beta\gamma}$, is defined by its commutator
\begin{equation}
T^{\alpha}_{\;\;\beta\gamma}=\Gamma^{\alpha}_{\;\;\gamma\beta}-\Gamma^{\alpha}_{\;\;\beta \gamma}= h_{a}^{\;\;\alpha}\left(\partial_{\beta}h^{a}_{\;\gamma}-\partial_{\gamma}h^{a}_{\;\beta}\right)\,. \label{eq:nospintorsion}
\end{equation}
In this non-covariant version of the theory the torsion scalar is defined by
\begin{equation} 
    T :=
    \frac14  T_{\alpha\beta\gamma}  T^{\alpha\beta\gamma}
     + \frac12  T_{\alpha\beta\gamma}  T^{\gamma\beta\alpha}
     - T_{\alpha\beta}^{\;\;\;\alpha}T^{\gamma\beta}_{\;\;\;\;\gamma}\,, \label{eq:nospinT}
\end{equation}
and the action (\ref{eq:gravaction}) is varied with respect to the tetrad degrees of freedom to yield the extended teleparallel gravitational equations of motion
\begin{equation}
h^{-1} h^{a}_{\;\rho} \partial_{\mu} \Big( h \frac{F(T)}{dT} S_{a}^{\;\nu\mu} \Big)
- \frac{dF(T)}{dT} T_{\alpha\beta\rho} S^{\alpha\beta\nu}
+ \frac{1}{2} F(T) \delta_{\rho}^{\;\nu} = 8\pi \mathcal{T}_{\rho}^{\;\nu}\,. \label{eq:nonceoms}
\end{equation}
Here $\mathcal{T}_{\rho}^{\;\nu}$ is the usual stress-energy tensor and 
\begin{equation} 
    S_{\alpha\beta\gamma} :=  K_{\beta\gamma\alpha}
      + g_{\alpha\beta} \,  T_{\sigma\gamma}^{\;\;\;\sigma}
      - g_{\alpha\gamma} \,  T_{\sigma\beta}^{\;\;\;\sigma} \label{eq:stensor}
\end{equation}
with 
\begin{equation} 
    K_{\alpha\beta\gamma} :=
     \frac12 \left(  T_{\alpha\gamma\beta}
     +  T_{\beta\alpha\gamma} +  T_{\gamma\alpha\beta} \right)\,. \label{eq:ktensor}
\end{equation}
The tensors $S_{\alpha\beta\gamma}$ and $K_{\alpha\beta\gamma}$ are known as the modified torsion and contorsion tensors respectively. As long as the tetrad yields zero for the inertial spin connection (to be discussed), the tetrad is a ``good'' one and the above theory is robust. However, sometimes the criteria listed earlier for a good tetrad are not sufficient to yield a tetrad which makes this spin connection vanish. In such a case one is inadvertently introducing non-gravitational effects into the equations of motion. Historically this issue of choosing an acceptable tetrad has been a rather difficult one, as isolating the correct spin connection and formulating the correct, fully covariant, equations of motion is rather difficult. Therefore the non-covariant theory is generally used and the price to pay is that there is a chance one may accidentally introduce inertial effects, even if the tetrad chosen satisfies the criteria mentioned earlier. 

Recently there has been a breakthrough with the situation of covariance in $F(T)$ gravity. Kr\v{s}\v{s}\'{a}k and Saridakis have shown how to identify the correct inertial spin connection to use and how to incorporate it consistently in the equations of motion \cite{ref:covariant} to yield a fully (spacetime and orthonormal-Lorentz) covariant theory. Briefly, in \cite{ref:covariant} the curvatureless spin connection (\ref{eq:spincon}) to be chosen is constructed as follows:
\begin{itemize}
\item Choose any metric compatible tetrad, $h^{a}_{\;\mu}$.
\item In this tetrad ``turn off'' gravity by taking $G\rightarrow 0$. This resulting tetrad is called $\hr$.
\item Calculate the torsion tensor with $\hr$, including the inertial spin connection, and set it to zero. That is, form the following equation:
\begin{equation}
 T^{a}_{\;\;\mu\nu}(\hr,\omega^{a}_{\;\;b\sigma})=\partial_{\mu}h^{a}_{\mbox{\tiny{(r)}}\nu} - \partial_{\nu}h^{a}_{\mbox{\tiny{(r)}}\mu} + \omega^{a}_{\;\;b\mu}h^{b}_{\mbox{\tiny{(r)}}\nu} - \omega^{a}_{\;\;b\nu}h^{b}_{\mbox{\tiny{(r)}}\mu} = 0 \label{eq:torinertial}
\end{equation}
and solve for the components of $\omega^{a}_{\;\;b\sigma}$.
\item The above $\omega^{a}_{\;\;b\sigma}$ is the one to use in the covariant formulation of $F(T)$ gravity in the equations that follow.
\end{itemize}

With the appropriate spin connection chosen as above, the full torsion tensor, which is the commutator of the Weitzenb\"{o}ck connection now with non-zero spin connection, is calculated via 
 \begin{equation}
 T^{\alpha}_{\;\;\beta\gamma}=h_{a}^{\;\;\alpha}\left(\partial_{\beta}h^{a}_{\;\gamma}-\partial_{\gamma}h^{a}_{\;\beta}\right) + h_{a}^{\;\;\alpha} \omega^{a}_{\;\;b\beta}h^{b}_{\;\;\gamma} - h_{a}^{\;\;\alpha} \omega^{a}_{\;\;b\gamma}h^{b}_{\;\;\beta}  \label{eq:torsionproper}
 \end{equation}
 and the torsion scalar, the modified torsion tensor, and the contorsion tensor are all calculated as in (\ref{eq:nospinT}), (\ref{eq:stensor}) and (\ref{eq:ktensor}) using (\ref{eq:torsionproper}).
 
 The resulting equation of motion are \cite{ref:covariant}
 \begin{equation}
\Scale[0.95]{h^{-1} h^{a}_{\;\rho} \partial_{\mu} \Big( h \frac{F(T)}{dT} S_{a}^{\;\nu\mu} \Big)
- \frac{dF(T)}{dT} T_{\alpha\beta\rho} S^{\alpha\beta\nu}
+ \frac{1}{2} F(T) \delta_{\rho}^{\;\nu} + \frac{dF(T)}{dT} S_{a}^{\;\;\alpha\nu} h^{b}_{\;\;\rho}\omega^{a}_{\;\;b\alpha} = 8\pi \mathcal{T}_{\rho}^{\;\nu}}\,. \label{eq:eoms}
\end{equation}
(Some factors differ from \cite{ref:covariant} due to an overall multiplicative factor difference in our expression for $T$. Equations (\ref{eq:eoms}) were derived from scratch using our conventions to ensure that they are compatible with the quantities as defined here.) 
 
In section \ref{sec:vac} we perturbatively solve the vacuum version of the above equations of motion and obtain the correction to the Schwarzschild vacuum in $T+(\alpha/2)\;T^{2}$ gravity. We perturb about the Schwarzschild solution since it is known the even in the regime where gravity is not very weak, the Schwarzschild solution provides an excellent approximation to the spherically symmetric field. (In fact, perturbations about Schwarzschild are commonly used in general relativity studies to take into account deviations from spherical symmetry due to higher polar moments.) This allows us to not just limit our study to solar system observations but also in the arena where deviations from Minkowski spacetime may not be small but deviations from the Schwarzschild vacuum is still small, such as in the vicinity of neutron stars. In the non-covariant theory there are tetrads in the literature which yield the Schwarzschild or related Kottler vacuum as a solution to $F(T)$ gravity even for the case $F(T)\neq T$ \cite{ref:schw1}, \cite{ref:schw2}. Interesting non-Schwarzschild vacua have also been obtained in \cite{ref:schw3}. This is because, before the covariant theory, any tetrad compatible with criteria such as those listed on page \pageref{page:criteria} could be considered acceptable as it was difficult to discern what was a truly gravitational tetrad and which incorporated Lorentz frame effects. The tetrads leading to the Schwarzschild vacuum turn out to not be compatible with vanishing inertial spin connection derived via the solution to (\ref{eq:torinertial}), as required for the non-covariant theory to coincide with the covariant theory. Given the difficulty in finding the appropriate tetrad in the non-covariant theory, it is understandable that those tetrads were considered. It can be shown that in the covariant theory, the Schwarzschild solution is not a vacuum solution for $F(T)\neq T$ by plugging in the Schwarzschild solution into (\ref{eq:eoms}) (with any metric compatible tetrad now since all yield the same equations of motion in the covariant theory). 

In section \ref{sec:orbits} we study some properties of test-particle orbits in the vacuum and derive the perihelion shift of orbits. A comparison of this perihelion shift with the observed value and its uncertainty for Mercury and in the vicinity of neutron stars allows us to set an upper limit on the value of the magnitude of the non-linear torsion coupling, $\alpha$. The bounds on the coupling set here differ from those in previous studies (\cite{ref:solar1}-\cite{ref:solarend}) due to the fact that the current study deals with the Lorentz covariant version of the theory, which differs from the standard formulations of $F(T)$ theory when the spin connection is non-zero. The rotated tetrads  utilized in \cite{ref:solar3},\cite{ref:solarend} meet the set of criteria set out on page \pageref{page:criteria} and therefore produce reasonable physics. However, those rotated tetrads do not yield zero components for the  inertial spin connection, and hence the theories there differ from the Lorentz covariant theory studied here due to the fact that the torsion tensors differ. This results in slightly different torsion scalars and modified torsion tensors (\ref{eq:stensor}). One can see how the resulting equations of motion are affected by noting how these quantities enter the equations of motion, and so if these quantities differ, different equations of motion can ensue. We hasten to add that this is by no means an implication that previous works are incorrect, it is simply that they are working in a different theory than the Lorentz covariant one considered here. Interestingly, in the literature, rotated spherically symmetric tetrads which do yield zero spin connection components can be found in \cite{ref:covariant}, \cite{ref:tamgood}, \cite{ref:tampres}, and it can be checked that such rotated tetrads do produce equations of motion equivalent to the covariant version of $F(T)$ gravity.

Solar system tests turn out to provide a higher bound to the value of the nonlinear coupling, due to the extremely weak nature of torsion in this regime. Given the precision of solar system measurements it is worthwhile noting this bound. In an arena where torsion is stronger, but still very close to general relativity so that the Schwarzschild solution is expected to dominate, such as in the vicinity of neutron stars, a similar bound on the nonlinear coupling can be achieved for consistency, which we also study in section \ref{sec:orbits}. We choose \neut as the system to study since there is fairly accurate data on its orbital properties. It also has a mass ratio slightly more favorable to the approximations made here than many other such binaries for which good data is available. Finally, we make some concluding remarks in section \ref{sec:conc}.

\section{The spherically symmetric vacuum}\label{sec:vac}
In this work we consider the arena of static spherical symmetry. The metric's line element can therefore be written as
\begin{equation}
 ds^{2}=A^{2}(r) dt^{2}-B^{2}(r) dr^{2}-r^{2}d\theta^{2}-r^{2}\sin^{2}\theta\,d\varphi^{2}\,. \label{eq:sphereline}
\end{equation}
Since we are now working in a fully covariant theory, for the corresponding gravitational tetrad it is easiest to consider the diagonal one aligned with the coordinate axes
\begin{equation}
\left[h^{a}_{\;\mu}\right]_{\mbox{\tiny{diag}}}=\left( \begin{array}{cccc}
A(r) & 0 & 0 & 0\\
0 & B(r) & 0 & 0 \\
0 & 0 & r & 0 \\
0 & 0 & 0 & r\sin\theta  \end{array} \right)\,. \label{eq:diagtet}
\end{equation}
Note that in the non-covariant theory this would have been a ``bad'' tetrad unless $F(T)=T$. Using the prescription in the introduction, the non-zero spin connection components, $\omega^{a}_{\;\;b\sigma}$, to be used which are compatible with (\ref{eq:diagtet}) are
\begin{equation}
\omega^{\hat1\hat2}{}_2 = - \omega^{\hat2\hat1}{}_2 = 1, \quad
\omega^{\hat1\hat3}{}_3 = - \omega^{\hat3\hat1}{}_3 = \sin\theta, \quad
\omega^{\hat2\hat3}{}_3 = - \omega^{\hat3\hat2}{}_3 = \cos\theta
\,. \label{eq:ourspincon}
\end{equation}

Within this framework we consider the theory given by the Lagrangian density
\begin{equation}
 \mathcal{L}=T+\frac{\alpha}{2}T^{2}\,, \label{eq:ourlag}
\end{equation}
where, from (\ref{eq:nospinT}),
\begin{equation}
T = - \frac{ 2 ( B(r) - 1 )( A(r) - A(r) B(r) + 2 r A'(r) )
          }{ r^2 A(r) B(r)^2} \, .
\end{equation}
The term linear in $T$ in (\ref{eq:ourlag}) yields general relativity (GR) results and the non-linear term yields a torsion analog of Starobinsky gravity. (As briefly discussed in the introduction, this does \emph{not} yield the same theory as the corresponding $F(R)=R+(\mu_{0}/2)\, R^{2}$ theory.) If one considers Lagrangian densities power expandable in $T$ the $T^{2}$ contribution is arguably the most important contribution beyond GR when torsion is weak. It is therefore a natural theory to study. By using $A(r)=\sqrt(1-2M/r)=B^{-1}(r)$ in the equations of motion (\ref{eq:eoms}) one can verify that the Schwarzschild solution is no longer a vacuum solution for $F(T)=T+(\alpha/2)\,T^{2}$. 

We write the specific gravitational action we are dealing with in the following way
\begin{equation}
 S=\frac{1}{2\kappa}\int T\left(1+\varepsilon \frac{\alpha}{2} T\right) \mbox{det}[h^{a}_{\;\mu}]\,d^{4}x \,, \label{eq:ouraction}
\end{equation}
as we consider the theory we are dealing with as not necessarily the full theory, but approximately correct when nonlinear torsion effects (i.e. deviations from general relativity) are small. We make the mild assumption that the full $F(T)$, whatever it may be, is analytic in $T$, and hence (\ref{eq:ouraction}) captures the dominant terms in the Taylor expansion. It also may be that the $T^{2}$ contribution may be exact, and many authors have studied specific power-law Lagrangians as an exact theory \cite{ref:power1} - \cite{ref:power5}.

The equations of motion (\ref{eq:eoms}), even with $\mathcal{T}_{\rho}^{\;\nu}=0$ as required for vacuum, are extremely complicated to solve even when frozen to spherical symmetry. We therefore adopt a perturbative strategy and write the relevant tetrad components in (\ref{eq:diagtet}) as
\begin{equation}
 A(r)=\underset{\mbox{\tiny{$0$}}}A(r) + \varepsilon a(r),\quad B(r)=\underset{\mbox{\tiny{$0$}}}B(r) + \varepsilon b(r)\,. \label{eq:pertet}
\end{equation}
In (\ref{eq:ouraction}) and (\ref{eq:pertet}) the constant $\varepsilon$ is simply used to keep track of what is a small quantity. We will ignore effects of order greater than $\varepsilon$ and in the end this tracking constant will be set equal to 1. However, it is very important to check at all stages that quantities which multiply $\varepsilon$ are very small both in comparison to 1 as well as in comparison to the terms not multiplied by $\varepsilon$. This ensures both that the perturbative approach is valid as well as enforces the condition that the theory is in a regime where general relativity results ($F(T)=T$) are the dominant ones. This is why we restrict ourselves in section \ref{sec:orbits} to observations in the relatively weak field.

As we are considering the spherically symmetric vacuum in the weak torsion limit, we perturb the tetrad around the Schwarzschild value, which we know yields very good agreement with observations when gravity is weak. That is, we set
\begin{equation}
 \underset{\mbox{\tiny{$0$}}}A(r)=\underset{\mbox{\tiny{$0$}}}B^{-1}(r)=\sqrt{1-\frac{2M}{r}}\,, \label{eq:abzero}
\end{equation}
with the assumption that whatever the ``true'' vacuum solution is in the full theory, it must be close to the well tested Schwarzschild solution unless gravity is very strong.
It is now a matter of plugging in the tetrad components into the equations of motion (\ref{eq:eoms}) and solving, to order $\varepsilon$, the vacuum equations. The resulting equations can be simplified by writing them as nonautonomous equations of the function $\mu:=[1 - 2M/r]^{1/2}$, as well as of the functions to be solved for. These equations are:
\begin{subequations} \romansubs
{\allowdisplaybreaks\begin{align}
a'(r) & = \frac{ r^2 \mu ( 1 - \mu^2 ) a(r)
        + 2 r^2 \mu^3 b(r) -2 \alpha (1 - \mu)^4 }{ 2 r^3 \mu^3 }\,, \label{eq:aeq} \\[0.1cm]
b'(r) & = - \frac{ r^2 \mu^3 ( 3 - \mu^2 ) b(r)
        +2 \alpha ( 1 - \mu )^3 ( 1 + 5\mu + 10\mu^2 ) }{ 2 r^3 \mu^5 }\,. \label{eq:beq}
\end{align}}
\end{subequations}
Although complicated, surprisingly, they can be solved. The solutions can be written as:
\begin{subequations} \romansubs
{\allowdisplaybreaks\begin{align}
a(r) & = a_0 \mu - \frac{b_0}{r \mu} - \frac{ 51 - 93 \mu^2 - 128 \mu^3
       + 45 \mu^4 - 3 \mu^6 - 12(1-3\mu^2) \ln\mu
       }{ 6 r^2 \mu ( 1 - \mu^2 ) ^2 } \, \alpha \, ,\label{eq:asol} \\[0.1cm]
b(r) & = \frac{b_0}{r \mu^3}
       + \frac{ 63 - \mu ( 24 - \mu ( 12 + \mu ( 64 - 3 \mu ( 25 - 8 \mu ) ) ) )
       - 12 \ln\mu }{ 6 r^2 \mu^3 ( 1 - \mu^2 ) } \, \alpha \, ,\label{eq:bsol}
\end{align}}
\end{subequations}
The values of the integration constants, denoted $a_0$ and $b_0$,
can be obtained from the requirement that as $r\to\infty$
the metric remains asymptotically Schwarzschild,
with unchanged mass parameter, $M$. (This also ensures the correct Newtonian limit far away.)
Expanding the solutions in powers of $1/r$ gives
\begin{subequations}\romansubs
{\allowdisplaybreaks
\begin{align}
a(r)=& a_{0} + \frac{8\alpha}{3M^{2}}-\left(b_{0}
+ a_{0} M + \frac{16 \alpha}{3M}\right) \frac{1}{r}
+ \mathcal{O}\left(\frac{1}{r^{2}}\right)\,, \label{eq:aasympt}\\[0.1cm]
b(r)=& \left(b_{0} + \frac{8\alpha}{3M}\right) \frac{1}{r}
+ \mathcal{O}\left(\frac{1}{r^{2}}\right)\,. \label{eq:basympt}
\end{align}}
\end{subequations}
Setting $a_{0}={-8\alpha}/{3M^{2}}$
and setting $b_{0}={-8\alpha}/{3M}$
yields the appropriate Schwarzschild limit
with correct fall-off properties far from the gravitating object and does not spoil the Bondi-Metzner-Sachs group symmetry structure of asymptotic infinity.
With these values, the leading terms
in the power expansion of $a(r)$ and $b(r)$ are found to be
\begin{equation} \label{eq:absolseries}
a(r) = - \frac{2M^3\alpha}{5 r^5} + \mathcal{O}\left(\frac{M^{4}\alpha}{r^6}\right),
\qquad 
b(r) = \frac{2M^3\alpha}{r^5} + \mathcal{O}\left(\frac{M^{4}\alpha}{r^6}\right).
\end{equation}
It is also interesting to compute the leading term of the power expansion
of the quantity $\frac{\alpha}{2} T$ which is in our perturbation
scheme assumed to be a small quantity. For this we obtained
\begin{equation} \label{eq:alphaOverTwoTorsionSeries}
\frac{\alpha}{2} T
= - \frac{M^2 \alpha}{r^4} + \mathcal{O}\left(\frac{M^{3}\alpha}{r^5}\right) .
\end{equation}

\section{Observational limits on nonlinear torsion}\label{sec:orbits}
In this section we will use the solution derived in section \ref{sec:vac} to set limits on the magnitude of the nonlinear coupling $\alpha$. Determining this coupling constant is important since it governs the deviations from general relativity. Therefore knowing the value of this constant will allow one to rule out torsion effects in the universe if these effects require an $\alpha$ whose value is ruled out by observation (for example, attributing the current observed acceleration of the universe to nonlinear torsion effects, etc.) Since we are working in the weak field limit, a natural observational phenomenon to consider is the periastron shift of objects orbiting a more massive spherical object. Specifically we use the current uncertainty in the perihelion shift of Mercury to set a limit on the nonlinear coupling $\alpha$. Various bounds from solar system observations have been performed in the interesting works of \cite{ref:solar1} - \cite{ref:solarend} within the non-covariant theory. Some of these studies utilize the diagonal tetrad or other tetrad which has non-vanishing inertial spin connection, so the theory is different from the covariant one we consider here. Also, in this study we are considering perturbations about the Schwarzschild solution, not Minkowski spacetime. This allows us to push the analysis to a regime where the field is stronger than in the solar system, but still considered ``weak field'' in the sense that the Schwarzschild solution is known to yield good agreement with observation. From this we can perform a consistency check with the stronger field regime. In this vein we also consider the uncertainty in the periastron shift of the binary system \neut, where as discussed below, we expect the Schwarzschild result to yield at least a reasonably reliable result.

\subsection{Perihelion shift}\label{eq:ps}
Here we will outline briefly the perihelion shift calculation. The calculation begins with the four-velocity relation $u^{\mu}u_{\mu}=1$, which in this case we can write as follows:
\begin{equation} \label{eq:drdtau2}
\left( \frac{dr}{d\tau} \right)^2 = \frac1{B^2}
\left( \frac{ E^2}{A^2} - \frac{ L^2}{r^2} - 1 \right)
\end{equation}
Here the constants of the motion are $E:=u_{t}$ and $L:=-u_{\phi}$, and due to the symmetry of the problem we have set $\theta=\pi/2$ and $u^{\theta}=0$.
One can also write
\begin{equation} \label{eq:drdphi2}
\left( \frac{dr}{d\varphi} \right)^2 = \frac{r^2}{B^2}
\left( \frac{r^2  E^2}{ L^2 A^2} - \frac{r^{2}}{ L^2} -1 \right).
\end{equation}
Equations (\ref{eq:drdtau2}) and (\ref{eq:drdphi2}) hold in general,
while in case of a strictly circular orbit with $r=r_0$,
(\ref{eq:drdtau2}) allows $ E$ and $ L$
to be expressed in terms of the metric functions
and the assumed radius of the orbit as 
\begin{equation} \label{eq:r0tildes}
 L^2 = \frac{ r_0^3 A'(r_0) }{A(r_0) - r_0 A(r_0)}, \qquad
 E^2 = \frac{ A(r_0)^3 }{A(r_0) - r_0 A(r_0)}.
\end{equation}
In order to study the precession of close-to-circular orbits (for more eccentric orbits we make a minor modification below)
we write the radial coordinate as $r(\varphi)=r_0 + y(\varphi)$
and assuming $y/r_0$ is small
we expand (\ref{eq:drdphi2}) in powers of this small parameter. 
We obtain a differential equation of the form
$y'(\varphi)^2 = f_{0} + g_{0} y(\varphi) - k^2 y(\varphi)^2 + \mathcal{O}(y^3)$
and using (\ref{eq:r0tildes}) we find $f_{0}=g_{0}=0$ and
\begin{equation}
k^2 = \frac{1}{B(r_0)^2}
\left( 3 - \frac{ 3 r_0 A'(r_0) }{ A(r_0) }
         + \frac{ r_0 A''(r_0) }{ A'(r_0) } \right) .
\end{equation}
% We obtain
% $y'(\varphi)^2 = - k^2 y(\varphi)^2 + \mathcal{O}(y^3)$,
% where the quantity $k^2$ can be written
% with the help of (\ref{eq:r0tildes}) as
% \begin{equation}
% k^2 = \frac{1}{B(r_0)^2}
% \left( 3 - \frac{ 3 r_0 A'(r_0) }{ A(r_0) }
%          + \frac{ r_0 A''(r_0) }{ A'(r_0) } \right) .
% \end{equation}
The solution to the differential equation is
oscillatory with the `wave-number' $k$ which is related to the precession angle $\Delta\phi$ by
\begin{equation}
\Delta\phi_{\text{per cycle}} = 2\pi \left( \frac1k - 1 \right).
\end{equation}
This quantity is now computed with the metric functions
previously derived describing the vacuum solutions in covariant $F(T)$ theory.
A series expansion in the small parameters $M/r_{0}$ and $\varepsilon$ yields the result
\begin{equation}
 \Delta\phi =\left[ \frac{6\pi M}{r_{0}}+\frac{27\pi M^{2}}{r_{0}^{2}} + \mathcal{O}\left(\frac{M^{3}}{r_{0}^{3}}\right)\right] +
 \left[ \frac{8\pi\alpha}{r_{0}^{2}}\left(\frac{M^{2}}{r_{0}^{2}}\right) + \mathcal{O}\left(\frac{\alpha M^{3}}{r_{0}^{5}}\right)\right] \varepsilon + \mathcal{O}\left(\varepsilon^{2}\right)\,. \label{eq:perseries}
\end{equation}
The terms in the first square bracket yields the usual general relativity result. The largest contributions to planetary perihelion shift due to nonlinear torsion effects are the terms shown in the second square bracket. 

\begin{figure}[h!t]
\begin{center}
\includegraphics[scale=1.00, keepaspectratio=true]{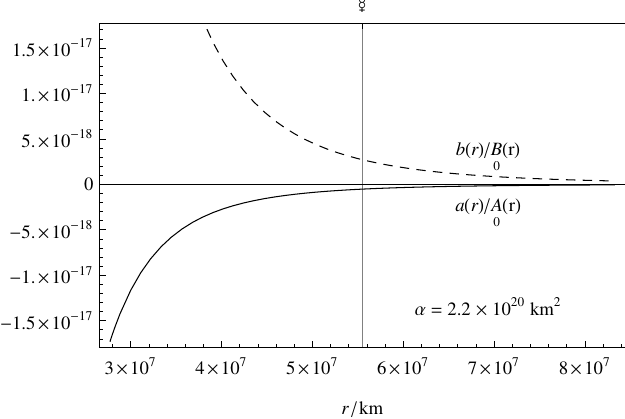}
\caption{{\small{A plot of $a(r)/\underset{\mbox{\tiny{$0$}}}A(r)$ (solid) and $b(r)/\underset{\mbox{\tiny{$0$}}}B(r)$ (dashed) in the vicinity of Mercury's orbit.}}}
\label{fig:mercsmallcheck}
\end{center}
\end{figure}

Before continuing it is important to confirm that we are in a regime where the corrections to the metric, (\ref{eq:asol}) and (\ref{eq:bsol}), are much smaller than the Schwarzschild values, $\underset{\mbox{\tiny{$0$}}}A(r)$ and $\underset{\mbox{\tiny{$0$}}}B(r)$. We also need to check that $(\alpha/2)\,T \ll 1$. For the solar system, in the vicinity of Mercury's orbit, we have $M=M_{\mbox{\tiny{$\astrosun$}}}=1.47$ km and $r_{0}=r_{\mbox{\tiny{$\mercury$}}}=5.55\times 10^{7}$km. In figure \ref{fig:mercsmallcheck} we plot $a(r)/\underset{\mbox{\tiny{$0$}}}A(r)$ and $b(r)/\underset{\mbox{\tiny{$0$}}}B(r)$ in the vicinity of Mercury's orbit which confirms that the deviation from the Schwarzschild solution is small. 

\begin{figure}[ht]
\begin{center}
\includegraphics[scale=1.00, keepaspectratio=true]{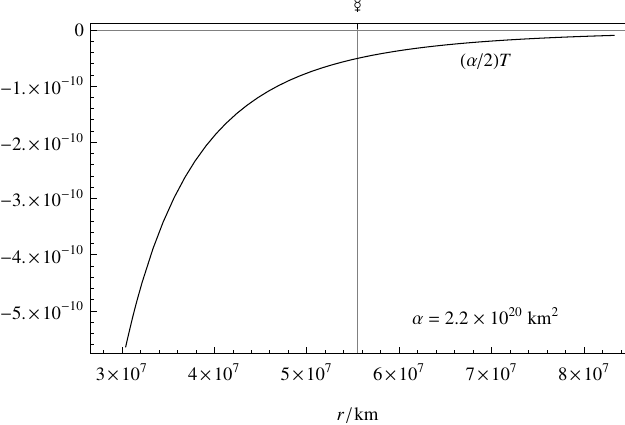}
\caption{{\small{The quantity $(\alpha_{\mbox{\tiny{max$\mercury$}}}/2) T$ in the vicinity of Mercury's orbit. Since this quantity is less than 1 the nonlinear torsion effects are subdominant to general relativity as required in the calculation presented here.}}}
\label{fig:nonlinsmall}
\end{center}
\end{figure}

We are looking to set an upper bound on the magnitude of $\alpha$ via solar system observations of Mercury's perihelion shift. For the perihelion precession there is the observed residual value of the precession (meaning the value that could not be accounted for before general relativity), and on the theoretical side there is the calculated value of the perihelion precession from general relativity. The observed value of Mercury's residual perihelion shift is $42.98''$/cy with an observational uncertainty of approximately $0.04''$/cy \cite{ref:mercper1}, \cite{ref:mercperend}. The prediction from the first two terms in (\ref{eq:perseries}) is $42.91''$/cy. To get an upper bound on the magnitude of $\alpha$ we make the gross assumption that the nonlinear torsion terms can contribute to this general relativity result and raise it to a maximum of $43.02''$/cy (the observational value plus its uncertainty). In other words, for an upper bound on $\alpha$ the nonlinear torsion effects can make up for the complete difference in the number calculated from the first square bracket in (\ref{eq:perseries}) and the maximum possible value for the perihelion precession. There are, of course, more general relativity contributions to the perihelion shift than those encoded in the first two terms in (\ref{eq:perseries}) which bring the general relativity result to within the experimental uncertainty (these contributions are due to effects from higher-order in $M/r_{0}$ in the first square bracket, general relativity corrections from the rotation of the sun, etc.) but, as we are interested on an upper bound on $\alpha$, we do not consider these. By considering the terms in the second square bracket in (\ref{eq:perseries}) we find that the value of $\alpha$ which brings the perihelion precession to its maximum possible value is
\begin{equation}
 |\alpha_{\mbox{\tiny{max$\mercury$}}}|=2.2 \times 10^{20}\,\mbox{km}^{2}\,. \label{eq:alphaper}
\end{equation}
Although this value is large, it should be noted that it is not a dimensionless quantity. It may be made arbitrarily large or small by a simple change of units. The fall-off of the contributions which arise from the $T^{2}$ term in the action are much faster than those of the TEGR term, hence allowing for a large value of the magnitude of $\alpha$ in these units in the vicinity of Mercury. Due to this fall-off behavior the allowed values of $|\alpha|$ are orders of magnitude larger than the ones obtained in Refs. \cite{ref:solar1}-\cite{ref:solarend} which span the order of $10^{-9}$ km$^{2}$ - $10$ km$^{2}$ for the $T^{2}$ couplings. As mentioned earlier, what is important for this bound to be meaningful is that the dimensionless quantity $\alpha T$ is much less than 1. In figure \ref{fig:nonlinsmall} we plot $(\alpha_{\mbox{\tiny{max$\mercury$}}}/2) T$ in the vicinity of Mercury's orbit and indeed, even with this value of $\alpha$, the nonlinear torsion term is subdominant to the TEGR term.

A quantity which is arguably equally interesting is the ratio of the nonlinear precession term to the general relativity precession term, which is formed as
\begin{equation}
 \left(\frac{\mbox{Nonlinear term}}{\mbox{GR term}}\right)=\frac{16M\alpha}{27Mr_{0}^{2}+6r_{0}^{3}}\,. \label{eq:mercratio}
\end{equation}
Using the mass of the Sun and mean radius of Mercury's orbit, (\ref{eq:mercratio}) yields a value of
\begin{equation}
\left( \frac{\text{Nonlinear term}}{\text{GR term}} \right)_{\mbox{\small{$\mercury$}}}
= \frac{1.15 \times 10^{-23}\alpha}{\mbox{km}^{2}}\,. \label{eq:mercratio2}
\end{equation}

We can set an independent bound on the nonlinear coupling in a stronger field regime, where the Schwarzschild solution is still deemed valid and any nonlinear torsion deviations are still small in comparison. The binary pulsar \neut provides such an arena. This binary consists of a pulsar of mass approximately $1.4 \msun$ orbiting a companion star of approximate mass $8.8 \msun$ \cite{ref:kaspi} with a period of 51 days. Although this mass ratio is not ideal, it does allow us to very roughly approximate the neutron star as a test particle in the gravitational field of the companion. (Many other such systems' mass ratios do not allow for this approximation, even roughly.) Since the eccentricity of the orbit is large, the radius $r_{0}$ in (\ref{eq:perseries}) is replaced by $a(1-\epsilon^{2})$ where $a$ is the semi-major axis (approximately $1.12\times 10^{7}$km) and $\epsilon$ is the eccentricity of the orbit (approximately 0.8) \cite{ref:kaspi1994}. It is noted that these approximations will make the \neut calculation less robust than  the corresponding one for Mercury, but it should provide an independent check in a stronger field regime. The final ingredient required to set the bound is the observational (from radio timing) value of the periastron shift which is $0.0259$ deg/yr with an uncertainty of approximately 0.0005 deg/yr \cite{ref:kaspi}. Note that this precession is much larger than Mercury's value.

If we perform the same analysis on \neut as with Mercury we obtain a bound of $|\alpha_{\mbox{\tiny{max \neut}}}|= 3\times10^{18}\mbox{km}^2$. Although smaller, this is arguably a less accurate bound than the Mercury one (due to the reasons discussed in the previous paragraph) but is not completely out of line with that result. This of course is due to the fact that we are treating the mass of the neutron star companion as small in comparison to the more massive star, and ignoring other effects such as distortions from sphericity of the star's material. A better modeling of the system in GR as a two-body system would yield quantities in the first bracket of (\ref{eq:perseries}) much closer to the actual precession, and hence provide a more accurate upper bound of $\alpha$. If we take Robertson's interesting result \cite{ref:robertson} that the mass $M$ in the GR leading term be replaced by the sum of the masses (from a low-order approximation to the two-body problem in GR), we obtain a slightly different value of  $\alpha_{\mbox{\tiny{max \neut}}}=2.1\times 10^{18}$km$^{2}$.

Again taking the first bracket in (\ref{eq:perseries}) to be the general relativity result (as with Mercury, spin effects and higher order $M/r$ terms will affect the GR prediction, but we again ignore this for an upper bound on $\alpha$), the leading nonlinear torsion term is $16\pi\alpha M^{2}/r_{0}^4$ and the leading GR terms are $6\pi M/ r_{0} + {27\pi M^{2}}/{r_{0}^{2}}$ . The ratio of these two terms is then
\begin{equation}
\left(\frac{\mbox{Nonlinear term}}{\mbox{GR term}}\right)_{|\mbox{\tiny{\neut}}}=\frac{1.8 \times 10^{-19}\alpha}{\mbox{km}^{2}}\,. \label{eq:neutratio}
\end{equation}

From the ratios (\ref{eq:mercratio2}) and (\ref{eq:neutratio}) it seems in both cases the observational bound of $\alpha$, taking into consideration the applicability issues of the approximations for \neut,  would be of the order $10^{19}\mbox{km}^{2}$. We re-stress that the largeness of this number is simply due to the units chosen.

\begin{figure}[h!t]
\begin{center}
\includegraphics[scale=1.00, keepaspectratio=true]{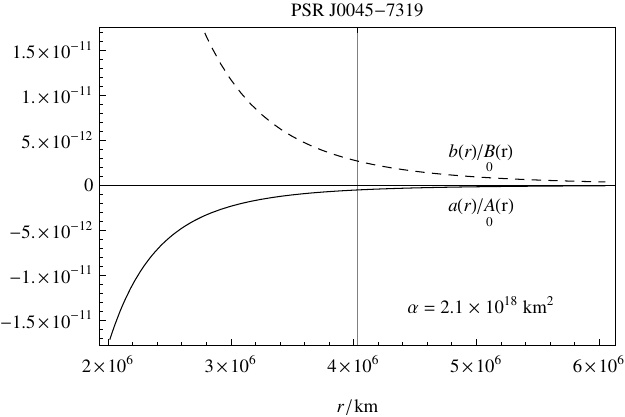}
\caption{{\small A plot of $a(r)/\underset{\mbox{\tiny{$0$}}}A(r)$ (solid) and $b(r)/\underset{\mbox{\tiny{$0$}}}B(r)$ (dashed) in the vicinity of $r=a(1-\epsilon^{2})$ for the binary orbit of \neut about its more massive companion star.}}
\label{fig:psrsmallcheck}
\end{center}
\end{figure}

\begin{figure}[h!t]
\begin{center}
\includegraphics[scale=1.00, keepaspectratio=true]{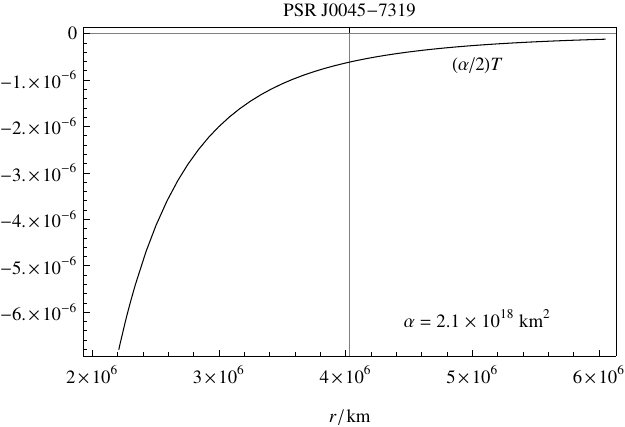}
\caption{{\small The quantity $(\alpha_{\mbox{\tiny{max \neut}}}/2) T$ in the vicinity of $r=a(1-\epsilon^{2})$ for the binary orbits of \neut about its more massive companion star. As with Mercury, this quantity is substantially less than 1 so that the nonlinear torsion effects are subdominant to general relativity as required in the calculation presented here.}}
\label{fig:nonlinpsrsmall}
\end{center}
\end{figure}

As before it is important to ensure that we are in a regime where our solution is valid so, in figures \ref{fig:psrsmallcheck} and \ref{fig:nonlinpsrsmall}, we plot similar parameters to the previous two figures (refer to figure caption). It can be seen that these values are small in the vicinity of our analysis.

\vspace{-0.3cm}

\section{{\small CONCLUDING REMARKS}}\label{sec:conc}
We have perturbatively solved for the spherically symmetric vacuum in the recently developed Lorentz covariant $F(T)$ gravity theory. The spherically symmetric vacuum is arguably one of the most interesting arenas of physical study in gravitational physics and therefore knowing its structure, even if at the perturbative level, is of relevance. Assuming that for small values of the torsion the first nonlinear term in the gravitational Lagrangian is of order $T^{2}$, we set a bound on the magnitude of the nonlinear coupling constant. The bound was set from considering the observational uncertainty in the perihelion shift of Mercury, as well as another check by considering the periastron precession of the binary system \neut. Both these calculations are consistent with an upper bound on the coupling, $\alpha$, of approximately $10^{20}$km$^{2}$. Any phenomena which one wishes to attribute to nonlinear torsion should therefore not require an $\alpha$ of greater value than this order, unless one is dealing in the strong torsion regime where $\mathcal{O}(T^{3})$ effects are expected to play an important role in the expansion of the action. It should be noted that this ``weak field'' regime actually encompasses a rather wide range of gravitational phenomena. The value of the torsion scalar $T$, even in the vicinity of \neut, is approximately of order $|T|\approx 10^{-24}$km$^{-2}$. Therefore it would require rather extreme gravitational conditions before one would be in what could be called the ``strong field'' regime. 

\section*{{\small ACKNOWLEDGMENTS}}
We are grateful to D. Horvat and Z. Naran\v{c}i\'{c} for helpful discussions regarding this project. This work is partially supported by the VIF2015 programme of the University of Zagreb. We thank C. Pfeifer, S. Bahamonde, and K. Flathmann for pointing out a typographical error in a previous version of this manuscript. (The results are unchanged due to the purely typographical nature of the issue.)

\vspace{0.5cm}
%%\newpage
\linespread{0.6}
\bibliographystyle{unsrt}

\end{document}